\documentclass[12pt,preprint]{aastex}

\newcommand{\bdv}[1]{\mbox{\boldmath$#1$}}

\def\au{{\rm au}} 
 
\def\kms{{\rm km}\,{\rm s}^{-1}}
\def\masyr{{\rm mas}\,{\rm yr}^{-1}}
\def\kpc{{\rm kpc}}

\def\max{{\rm max}}
\def\min{{\rm min}}
\def\rel{{\rm rel}}

\def\eff{{\rm eff}}
\def\rot{{\rm rot}}

\def\e{{\rm E}}
\def\bpi{{\bdv\pi}}
\def\bmu{{\bdv\mu}}

\begin{document}
\title{GEO and LEO: The Final Frontier for Plutonic FFP Parallax}

\author{\textsc{
Andrew Gould$^{1,2}$
}}

\affil{$^{1}$Department of Astronomy, Ohio State University, 140 W.
18th Ave., Columbus, OH 43210, USA}

\affil{$^{2}$Max-Planck-Institute for Astronomy, K\"{o}nigstuhl 17,
69117 Heidelberg, Germany}

\begin{abstract}

  I show that microlens parallaxes, $\bpi_\e$, can be derived for
  free-floating planets (FFPs) with masses down to that of Pluto,
  by combining observations from a satellite in geosynchronous (GEO)
  orbit with another observatory that is on or near Earth's surface,
  i.e., either ground-based or in low Earth orbit (LEO).  Because these
  low-mass FFPs typically have measurements of the angular
  Einstein radius, $\theta_\e$,
  from finite-source effects, such $\bpi_\e$ measurements directly yield
  the FFP mass $M=\theta_\e/\kappa \pi_\e$ where $\kappa$ is a physical
  constant.  Such Earth-GEO measurements almost perfectly complement
  Earth-L2 measurements, which extend to higher FFP mass and greater
  FFP distance.  LEO-only observations can yield mass measurement
  at even smaller FFP mass and nearer FFP distances.  I discuss methods
  for breaking the \citet{refsdal66} two-fold degeneracy in $\pi_{\e,\pm}$.

\end{abstract}


\section{{Introduction}
\label{sec:intro}}

It is at least plausible that there is a vast population 
of sub-planetary objects that can generate free-floating planet (FFP) type
signals, i.e., brief, single-lens single-source (1L1S) microlensing events
that show no trace of a host.  Both \citet{gould22} and \citet{sumi23}
derived FFP mass functions that were consistent with
$dN/d\log M \propto M^{-1}$, i.e., equal mass of planetary material
per logarithmic bin, albeit with large statistical uncertainties.
In particular, \citet{gould22} noted that if this power law were
extended by 18 orders of magnitude, it would also be consistent with
observations of interstellar asteroids and comets.

While FFPs may be truly ``free'' (i.e., not bound to any star), they
may also simply be so far from their hosts that these hosts do not
leave any trace on the event.  \citet{gould16} discussed how
to distinguish these two classes, and indeed how to further divide
the bound FFPs into various subclasses, and \citet{gould24}
examined this issue in greater detail.  In particular, \citet{gould24}
pointed out that if the host can be identified, then the FFP mass can
be measured using a slight variant of the method originally proposed
by \citet{refsdal64}:
\begin{equation}
  M = {\theta_\e^2\over \kappa\pi_\rel} = {(\mu_\rel t_\e)^2\over \kappa\pi_\rel};
  \qquad\kappa \equiv {4G\over c^2\au}= 8144\,{\mu{\rm as}\over M_\odot},
  \label{eqn:refsdal64}
\end{equation}
where $(\pi_\rel,\mu_\rel)$ are the lens-source relative
(parallax, proper motion) and $(\theta_\e,t_\e)$ are the
Einstein (angular radius, timescale),
\begin{equation}
  \theta_\e\equiv \sqrt{\kappa M \pi_\rel};\qquad
  t_\e\equiv{\theta_\e\over \mu_\rel}.
  \label{eqn:thetae-te}
\end{equation}
That is, once the host is identified, the host-source relative parallax
can be measured, and this is the same as the lens-source relative parallax,
$\pi_\rel$.  Moreover, even if $\theta_\e$ is not measured from
the event, it can be determined from two epochs of host-source
astrometry, which yield the host-source relative proper motion, which
is essentially the same as lens-source relative proper motion,
$\mu_\rel$.  Because $t_\e$ is measured during the event, this immediately
yields $\theta_\e=\mu_\rel t_\e$.

On the other hand if the FFP is unbound (or if the host cannot be identified),
then the only way to determine the mass is to measure the microlens
parallax,
\begin{equation}
  \bpi_\e\equiv \pi_\e {\bmu_\rel\over\mu_\rel};\qquad
  \pi_\e\equiv {\pi_\rel\over\theta_\e} = \sqrt{\pi_\rel\over\kappa M}.
  \label{eqn:piedef}
\end{equation}
Then,
\begin{equation}
  M = {\theta_\e\over\kappa\pi_\e}.
  \label{eqn:meval}
\end{equation}

Moreover, if the FFP is in the bulge, then (even if it is bound to an
identified host) the \citet{refsdal64} method of
Equation~(\ref{eqn:refsdal64}) will likely lead to a very uncertain
mass estimate because $\pi_\rel=\pi_l - \pi_s$ is the difference
of two very similar quantities.

Thus, obtaining microlens parallax measurements for FFPs is crucial
for understanding this population.

The main focus for FFP microlens parallax measurements has been on L2
parallaxes, whereby the event is observed simultaneously from the
ground and L2 \citep{gould03,zhu16,gould21,ge22}, from two satellites
in different L2 orbits \citep{euclid19,ban20,gould21,bachelet22}
or from L2 and low-Earth-orbit (LEO, \citealt{yan22}).  There are two
main reasons for this focus on L2, one theoretical and the other practical.

The theoretical reason is that to successfully employ the two-observatory
method to measure parallax that was originally proposed by
\citet{refsdal66}, the magnitude of the projected separation
of the two observatories, $D_\perp = |{\bf D}_\perp|$, must be approximately
matched to the projected Einstein radius of the target population,
\begin{equation}
  {\tilde r}_\e \equiv {\au\over \pi_\e}=\au\sqrt{\kappa M\over\pi_\rel}=
0.02\,\au\biggl({M \over M_\oplus }\biggr)^{1/2}
  \biggl({\pi_\rel \over 60\,\mu\rm as}\biggr)^{-1/2},
  \label{eqn:retilde}
\end{equation}
In the \citet{refsdal66} method, the \citet{pac86} parameters $(t_0,u_0,t_\e)$
of the event from each of the two observatories are measured,
and $\bpi_\e$ is inferred by
\begin{equation}
  \bpi_\e = (\pi_{\e,\parallel},\pi_{\e,\perp})
  = {\au\over D_\perp}(\Delta \tau,\Delta \beta);\qquad
  \Delta \tau \equiv {t_{0,1}-t_{0,2}\over t_\e}; \qquad
  \Delta \beta \equiv u_{0,1}-u_{0,2}.
  \label{eqn:refpar}
\end{equation}
Thus, if $D_\perp\ll {\tilde r}_\e$,
then the two observatories will see essentially the same event, while
if $D_\perp\gg {\tilde r}_\e$, then one of the two observatories will
not see the event at all.  \citet{gould24} argued that at a given
$D_\perp$, one could approximately probe the range of ${\tilde r}_\e$ satisfying
\begin{equation}
0.05\la {D_\perp\over{\tilde r}_\e}\la 2.
  \label{eqn:parlims}
\end{equation}
In their Figure~4, they showed that
this factor $\sim 40$ range in ${\tilde r}_\e$ (factor $40^2=1600$
in mass) is approximately centered on $M_\oplus$ for
L2 (i.e., $D_\perp\simeq 0.01\,\au$) separations and for the typical
range of $\pi_\rel$ that characterizes microlensing events seen toward
the Galactic bulge.  See also Figure~\ref{fig:par2}, below.
As $M_\oplus$ FFPs have been the first focus of interest
for next-generation microlensing experiments, L2 separations appear to
be ideal.

The practical reason is that the {\it Roman} and {\it Euclid} survey
telescopes will operate (or already are operating) in L2 orbits.
{\it Roman} will have a very large microlensing component, while {\it Euclid}
may adopt one later in its mission.  In addition, the {\it Earth 2.0 (ET)}
mission is specifically designed to acquire Earth-L2 parallax measurements.

The main point of \citet{gould24} was that by adjusting its observation
schedule, {\it Roman} could be made much more sensitive to Pluto-mass
FFPs, i.e., $M\sim 0.002\,M_\oplus$.  Although these are individually
more difficult to detect than Earth-mass FFPs, they could plausibly
be $\sim 500$ times more numerous, and hence a radically different FFP
population could be effectively probed.  However, their same Figure~4,
which showed that L2 separations are extremely well-matched to Earth-mass
FFP parallaxes, also shows (unsurprisingly) that they are poorly matched
to Pluto-mass FFP parallaxes.  In particular, for Plutos with sufficiently
large $\pi_\rel\ga 60\,\mu$as to be detected at all, the observatories are
too far apart to both observe the same FFP event.  While it would still be of
great interest just to detect these Plutos, this mismatch raises the question of
whether it is possible to measure the $\bpi_\e$ for this potentially very
numerous class of objects.

Here I discuss two approaches to tackling this challenge.

\section{{Two Channels of Microlens Parallax}
  \label{sec:two-channels}}

\citet{refsdal66} and \citet{gould92} charted two logically distinct channels
to measure $\bpi_\e$, i.e., simultaneous observation from two platforms, and
observation from a single accelerated platform, respectively.  In the
original \citet{refsdal66} approach, the two platforms were the Earth
and a satellite in solar orbit, so a baseline $D_\perp\sim {\cal O}(\au)$.
However, \citet{holz96} proposed to apply the same two-observatory concept on an
$\epsilon_{\rm TP} \sim 1/23$,500 times smaller scale by placing both
observatories
on Earth (``terrestrial parallax'', TP), while \citet{gould03} proposed to
apply it on an $\epsilon_{\rm L2} \sim 1/100$ smaller scale by
placing the observatories
on Earth and at L2, as described in Section~\ref{sec:intro}.  Given the
${\tilde r}_\e\propto M^{1/2}$ scaling of Equation~(\ref{eqn:retilde}),
one might expect these to have peak sensitivities that are smaller than
the original \citet{refsdal66} proposal by factors of
$\epsilon_{\rm TP}^2 = 2\times 10^{-9}$ and $\epsilon_{\rm L2}^2 = 10^{-4}$,
respectively.  Because the peak sensitivity\footnote{For this purpose,
I define the ``peak'' according to the geometric means of the limits
of Equation~(\ref{eqn:parlims}), i.e., the point at which
${\tilde r}_\e= D_\perp\sqrt{10}$, and evaluated at the mid-point of
the abscissa of Figure~4 from \citet{gould24}, i.e.,
$\log(\pi_\rel/\mu{\rm as})=1.5$,
or $\pi_\rel = 32\,\mu$as.  Substituting $D_\perp=1\,\au$,
Equation~(\ref{eqn:retilde}) then yields
$M = 0.04\,M_\odot$.} for a satellite at $1\,\au$
is actually $M_{\rm peak,\au}\sim 0.04\,M_\odot$, these scaling relations
predict
$M_{\rm peak,TP}=\epsilon_{\rm TP}^2 M_{\rm peak,\au}\sim 3\times 10^{-5} M_\oplus$ and
$M_{\rm peak,L2}=\epsilon_{\rm L2}^2 M_{\rm peak,\au}\sim 1.3\, M_\oplus$, respectively.
Comparing to the discussion in Section~\ref{sec:intro}, we see that this naive
reasoning accurately predicts the application that has been envisaged for
L2.  However, it seems to predict that the utility of TP will be centered
on unobservably small objects.  In fact, the two TP events observed
to date \citep{ob07224,ob08279} have been much more massive, but this
is because they were extreme magnification events,
as predicted by \citet{gould97}.

The second channel is to observe the event from a single observatory on
an accelerated platform.  The first such platform was Earth itself
\citep{gould92}, which yields so-called ``annual parallax'' (AP), but this
was soon followed by ``low-Earth orbit (LEO) parallax''
\citep{honma99} and
``geosynchronous orbit  parallax'' (GEO, \citealt{gould13}).

The spatial scale of LEO is the same as TP, i.e., $R_\oplus$, and thus
(just as in TP) the practical application presented by \citet{honma99}
relied on sharply peaked magnification structures, in this case, caustics,
to have a measurable effect.

The spatial scale and time scale of GEO are $a_{\rm GEO}\sim 6.6\,R_\oplus$ and
$P_{\rm GEO} =1\,$d, which would seem
to be ill-matched to the requirements of acceleration-based parallax
measurements.  To illustrate their feasibility, \citet{gould13}, chose
an event with a typical Einstein timescale, $t_\e=20\,$d, but an
effective timescale matched to the orbital period,
$t_\eff\equiv u_0 t_\e=P_{\rm GEO}$, i.e., with $u_0=0.05$.  In this way, the
parallax displacement was amplified by $A_\max=20$ over an orbital period, to
$\delta \ln A \sim \delta u/u_0 \sim \pi_\e A_\max a_{\rm GEO}/\au
= \pi_\e 20(2.8\times 10^{-4})\rightarrow 8 \times 10^{-4}$, where
 $\pi_\e\sim 0.14$ was the value chosen by \citet{gould13} for the illustration.
Nevertheless, \citet{gould13} argued that
these small oscillations could be reliably measured, in part because
they repeated over several days, and in part because there would be a rigorous,
and very precise, check from AP
on one of the two $\bpi_\e$ components
that were derived from these oscillations.  See his Figures~2 and 3.

Subsequently, \citet{mogavero16}, showed that GEO observations were well
matched to the task of measuring $\bpi_\e$ for $t_\e\sim P_{\rm GEO}=1\,$d,
i.e., Jupiter-class FFPs and very low mass brown dwarfs.

These examples illustrate that while the overall scaling relations
provide a useful guide, the concrete circumstances of each application
must be closely considered.

\section{{New Application of GEO to Dwarf-Planet Parallaxes}
\label{sec:geop}}

Pluto is 456 times less massive than Earth.  Hence, the peak sensitivity
for Pluto-mass objects should be achieved at projected separations, $D_\perp$
that are $\sqrt{456}\simeq 21.4$ times smaller than for Earth-mass objects,
i.e., $D_\perp = 0.01\,\au/21.4 = 11.0\,R_\oplus$, just 1.7 times larger
than a GEO orbit.  Given that the mass sensitivity spans several decades
centered on this peak, this factor 1.7 is quite minor.  The first incarnation
of this idea, i.e., separate observatories on Earth and in GEO, 
is qualitatively different from the GEO approach proposed by \citet{gould13}
and further explored by \citet{mogavero16}, which made use of a
single accelerated observatory at GEO and which
was discussed in Section~\ref{sec:two-channels}.

\begin{figure}
\plotone{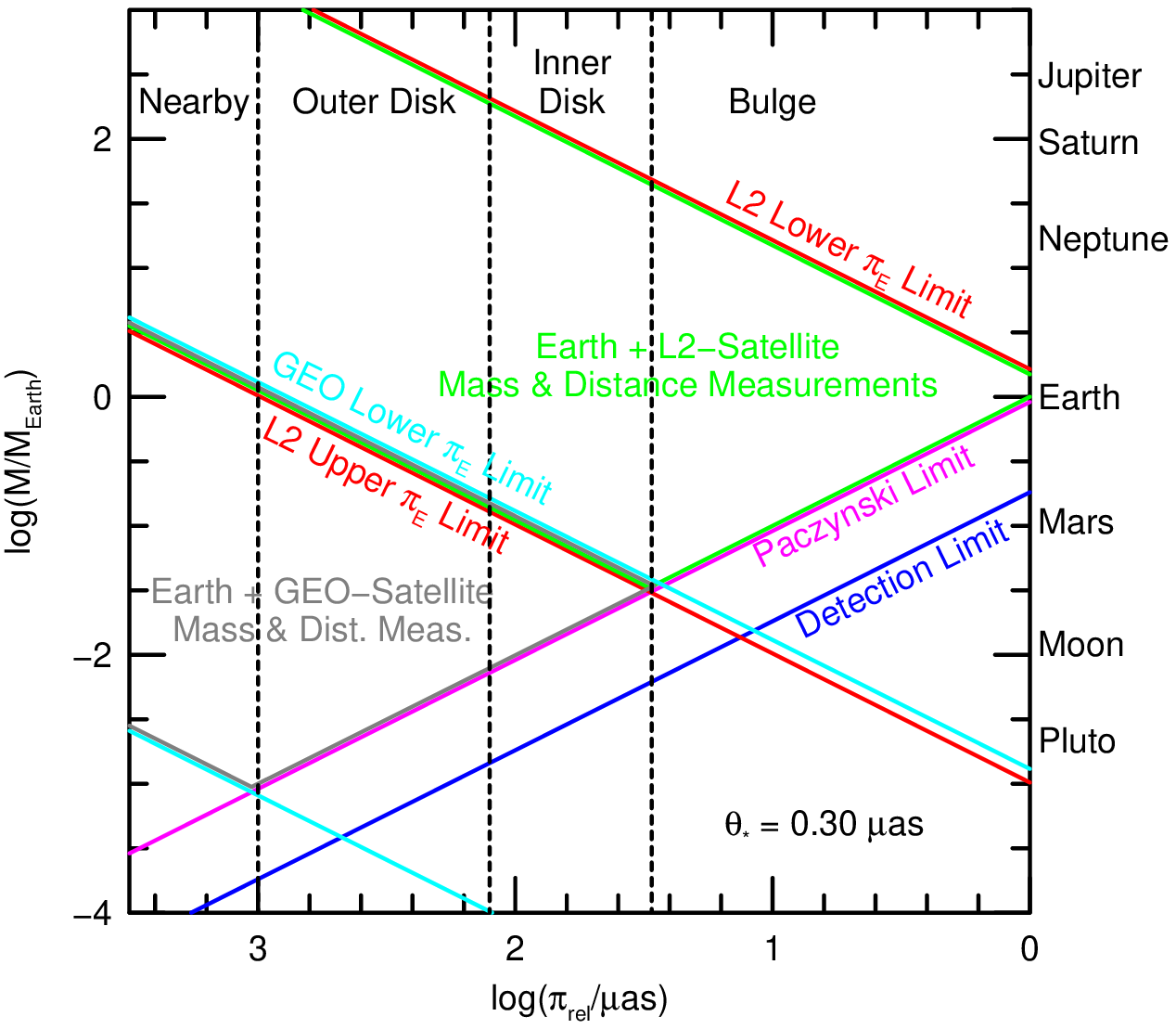}
\caption{Extended version of Figure~4 from \citet{gould24}, which
  focused on parallax measurements derived from two observatories
  separated by the Earth-L2 distance ($0.01\,\au$).  The red, green,
  magenta and blue lines are identical to that Figure.  The cyan and
  gray lines are new.  These illustrate the regions of
  parallax-measurement sensitivity based on observatories separated
  by the Earth-GEO distance on the FFP-mass/$\pi_\rel$ plane.
  Above the top cyan line, the two observatories see essentially the
  same event, while below the bottom cyan line, one observatory will
  not see the event at all.  Below the magenta line, the \citet{pac86}
  parameters cannot be measured, which undermines the $\bpi_\e$ measurement
  according to Equation~(\ref{eqn:refpar}).  $\bpi_\e$ measurements
  can therefore be made within the gray contours.  These reach to
  substantially lower FFP masses compared to Earth-L2 observations,
  but are restricted to the Galactic-disk FFPs (vertical lines).
    }
\label{fig:par2}
\end{figure}

I illustrate the sensitivity of this approach in Figure~\ref{fig:par2},
which is a version of Figure~4
from \citet{gould24} that has been extended to include GEO
parallaxes in addition to L2 parallaxes.  The gray lines show the
parallax limits of Equation~(\ref{eqn:parlims}) as applied to GEO
($D_\perp=6.63\,R_\oplus$), just as the red lines show these limits as applied to
L2 ($D_\perp=0.01\,\au$).  The magenta (``Paczy\'nski Limit'') and
blue (``Detection Limit'') lines are exactly the same as in Figure~4
of \citet{gould24}.  The Paczy\'nski Limit requires that
$\rho\equiv\theta_*/\theta_\e<2$ so that the \citet{pac86} parameters,
$u_0$ and $t_\e$, which are needed for Equation~(\ref{eqn:refpar}),
can be measured.  \citet{johnson22} showed that this becomes
increasingly difficult for $\rho\gg1$.  In particular, the gray
contours outline the region of $M$-$\pi_\rel$ space that is accessible
to GEO parallaxes, just as the green contours outline the region of
$M$-$\pi_\rel$ space that is accessible to L2 parallaxes.  Notably,
the two regions slightly overlap, so that their union covers a vast
swath of this space down to the Paczy\'nski limit.  One observes that
Moon-mass objects are accessible to GEO parallaxes for the
``outer disk'', i.e., $D_L\la 5\,\kpc$, while Pluto-mass objects are
accessible for the ``nearby'' population,
$D_L\la 1.3\,\kpc$.  By comparison, such objects are completely
inaccessible to L2 parallaxes, regardless of lens distance.

However, Figure~\ref{fig:par2} contains a hidden assumption, namely
that early M-dwarf sources, with $\theta_*=0.3\,\mu$as are observable.
That is, the Paczy\'nski limit shown in Figure~\ref{fig:par2} is based
on $\theta_*=0.3\,\mu$as, as being ``typical'' of {\it Roman} sources.
But for parallax to be measured, the same sources must be accessible
to the other observatory.  \citet{gould24} considered various
possibilities for the second observatory,
such as {\it Euclid} (also in L2), {\it CSST} in LEO,
as well as observatories on Earth.  Their Figure~4, showed only the
most favorable case for L2 parallaxes.  Regarding GEO parallaxes,
perhaps such early M-dwarf sources could be reached with LSST, and so
Figure~\ref{fig:par2} could be considered relevant to them.  For
example, the GEO satellite could be placed over the Western Pacific,
so that $D_\perp$ could be near its maximum value when the Galactic
bulge is near zenith in Chile.  Then, other, less powerful
observatories could be placed in Africa, which would take advantage of
complementary diurnal time periods.  Note that because the bulge fields have
Declination, $\delta\sim -29^\circ$, while GEO orbits are on the equator,
the minimum value of $D_\perp$ during the daily GEO cycle is
approximately $D_{\perp,\min}\sim a_{\rm GEO}\sin|\delta|\sim a_{\rm GEO}/2$,
depending on the precise position of the observatory on Earth.

\begin{figure}
\plotone{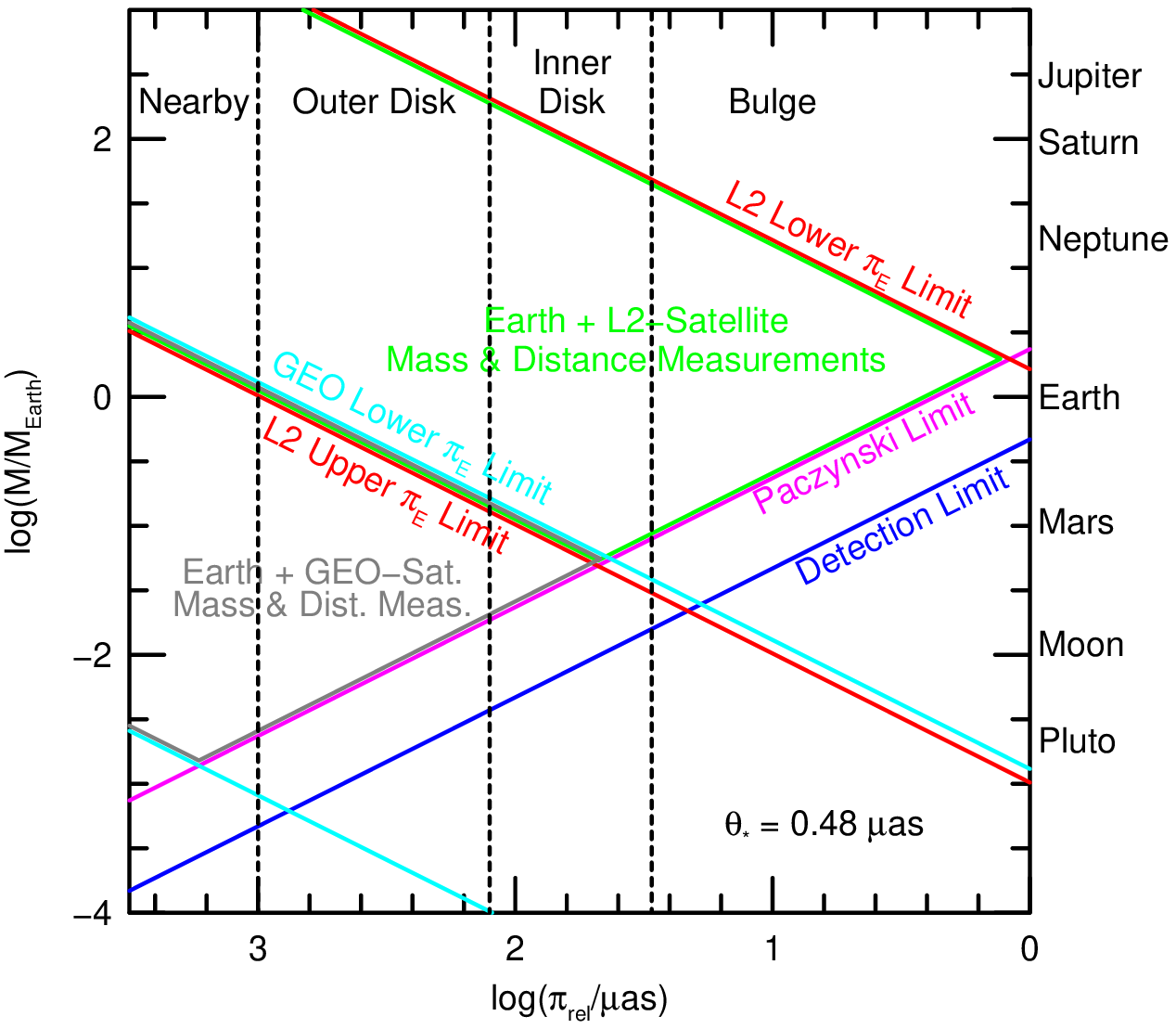}
\caption{Same as Figure~\ref{fig:par2} except that the blue and magenta
  lines are drawn under the assumption of an early K-dwarf source
  ($\theta_*=0.48\,\mu$as) rather than an early M-dwarf source
  ($\theta_*=0.30\,\mu$as).  The regions of sensitivity of both Earth-L2
  and Earth-GEO observations are reduced compared to Figure~\ref{fig:par2}.
  Figure~\ref{fig:par2} is more appropriate if the ``Earth'' observatory
  is a spacecraft in LEO orbit or LSST on the ground, while this Figure
  is more appropriate for typical ground-based survey telescopes.
    }
\label{fig:par3}
\end{figure}

Whether or not LSST could be convinced to undertake such dedicated
observations (see \citealt{gouldlsst}), one therefore should also consider
less capable observatories, i.e., either working to complement LSST
(as above) or instead of it.  To this end, I show in Figure~\ref{fig:par3},
a second view of the same parameter space, but adopting $\theta_*=0.48\,\mu$as,
which would be more accessible from typical ground-based surveys.
One sees that regions of the Galaxy in which Moon-mass and Pluto-mass
objects are detectable contracts significantly, although both
are still detectable at some lens distances.  Mathematically, the lower limit
on $\pi_\rel$ (at fixed $M$) increases by a factor $(0.48/0.30)^2 = 2.56$
between the two diagrams, which is $\sim 0.4$ dex, i.e., 2 tic marks
on the abscissa.

Another possibility for the second observatory would be a satellite in LEO.
This would have the advantage of space quality photometry and would yield
essentially the same $D_\perp$ as the ground-based observatories discussed above.
Such a LEO survey telescope, namely the {\it Chinese Space Station Telescope
  (CSST)} is already being planned by the National Astronomical Observatories
of China.  Compared to Earth-bound telescopes, a LEO observatory has
the important advantage of permanent good weather.  It can also observe
the bulge for a greater fraction of the 8760 hours of the year, being
only restricted by the horizon and the time that Sun is not in/near the
bulge\footnote{Depending on the phase of the 19-year Metonic cycle and the
baffling of the telescope, the Moon may interfere with observations as well.
This will also be more severe for ground-based than LEO observatories.},
rather than scattered sunlight in Earth's atmosphere.

A more ambitious option would be to place two telescopes in GEO, for example,
separated in phase by $180^\circ$.  In this case $D_\perp$ would oscillate
between $2\,a_{\rm GEO}$ and $2\sin|\delta| a_{\rm GEO}= 0.97\,a_{\rm GEO}$
over every 12 hours.  Observations could be continuous except when the
Sun was in or near the Galactic bulge.  That is, even when Earth and
the bulge were most closely aligned, the line of sight toward
the bulge would pass $(\sin\delta) a_{\rm GEO} - R_\oplus\simeq 2.21\,R_\oplus$
from Earth's surface.  Or, expressed in terms of angle,
$\psi = |\delta| - \sin^{-1}(R_\oplus/a_{\rm GEO}) \simeq 20.3^\circ$.

We can use Figure~\ref{fig:par2} to gauge the sensitivity to low-mass
planets and dwarf planets.  At minimum separation, $D_\perp\simeq a_{\rm GEO}$,
which is almost exactly what is represented in Figure~\ref{fig:par2}.
At maximum
separation, $D_\perp=2\,a_{\rm GEO}$, so that the cyan lines would be
moved upward by a factor 4 (i.e., 0.6 dex, slightly more than 1 tic mark)
relative to Figure~\ref{fig:par2}.  Hence, considering all phases of the
orbit, there would be a broad range of sensitivity to planets from
Earth-mass to Pluto-mass.

\section{{How to Break the Refsdal Parallax Degeneracy}
  \label{sec:degen}}

As already pointed out by \citet{refsdal66} (see also Figure~1
of \citealt{gould94}), Equation~(\ref{eqn:refpar}) does not lead to
a unique solution because $u_0$ is a signed quantity, but generally
only its amplitude can be derived from a microlensing light curve.
And this is particularly the case for very short events, which are the
sole object of interest in the present context.

This issue is of exceptional importance for low-mass FFPs because the
main statistical argument, the so-called ``Rich argument''
for breaking this degeneracy may not apply to them.  I explain both the
argument and its potential problems via a concrete example.

Suppose that $D_{\perp} = 6\,R_\oplus$, $\theta_*=0.3\,\mu$as, $\rho=1$,
$u_{0,1}=0.35$, $u_{0,2}=0.4$, $t_\e=0.66\,$hr, $\Delta \tau= 0.05$.
I first note that, independent of any issue related to parallax,
$\theta_\e = \theta_*/\rho=0.3\,\mu$as, and
$\mu_\rel = \theta_\e/t_\e= 4.0\,\masyr$.  Now, according to
Equation~(\ref{eqn:refpar}), there are four possible values
of $\Delta \beta$, i.e., $\pm(0.4-0.35)=\pm 0.05$ and
$\pm(0.4+0.35)=\pm 0.75$.  Therefore, there are two possible values
of $\sqrt{(\Delta \tau)^2 + (\Delta \beta)^2}$, namely
$\sqrt{0.05^2 + 0.05^2} \simeq 0.071$ and
$\sqrt{0.05^2 + 0.75^2} \simeq 0.75$, and therefore two values of
the parallax, $\pi_{\e,-} = 279$ and $\pi_{\e,+} = 2950$.  These
then lead to two pairs of values for $(M,\pi_\rel)$, namely
$(M/M_\oplus,\pi_\rel/\mu{\rm as})_- = (0.044,84)$ and
$(M/M_\oplus,\pi_\rel/\mu{\rm as})_+ = (0.0042,885)$.  These correspond to
a Mercury-mass object at $D_L\sim 5\,\kpc$ and a 2-Pluto-mass object
at $D_L\sim 1\,\kpc$.  According to the formal statement of the Rich argument
\citep{gould20}, the first solution is strongly favored by several factors,
including that there is 125 times more volume at 5 times the distance,
as well some Jacobian factors.  Therefore, in typical circumstances,
one would just ``choose'' the $\pi_{\e,-}$ solution as being overwhelmingly
probable.  Of course, this would lead to occasional mistakes, but these
would not affect any overall statistical conclusions.

However, in the present case, it could be that Pluto-class objects are
10 or even 100 times more common that Mercury-class objects.  If the
relative frequency were known to be very high, then this procedure
might lead to an ambivalent conclusion about this particular object,
despite the strong support for the Mercury-mass conclusion from the
geometric components of the Rich argument.  However, the real issue is
that we do not know this relative frequency a priori, and we have no
means to determine it except from this, and similar, experiments.
Therefore it is important to try to resolve the \citet{refsdal66}
degeneracy using data and not simply statistical arguments.

Two of the methods described in Section~\ref{sec:geop} provide paths
toward such resolution.

If the second observatory is in LEO, then there will
be a very strong signal for the 2-Pluto-mass solution because
${\tilde r}_{\e,+} = \au/\pi_{\e,+} = 8\,R_\oplus$, i.e., just 8 times
larger than the orbit.  On the other hand, for the Mercury-mass solution,
${\tilde r}_{\e,-}=85\,R_\oplus$, so the parallax signal would be unmeasurably small.
Note that in the example given, $u$ remains $u<1.5$ for an interval
$\delta t = 2.89\,t_\e=1.9\,$hr, (substantially more than the 90 min
orbital period), during which time the magnification, $A$,
remains $A(u)=(u^2+2)/u\sqrt{u^2+4}>17/15$.  Therefore, the parallax signal
is clearly detectable despite the roughly 50 min gap in the data due
to Earth occultation.  The signal becomes more difficult to recover for
cases that the magnification does not remain significant on both sides
of an Earth occultation.

If the second observatory is also in GEO, then this opens the possibility
of the ``3-observatory'' method of resolving the \citet{refsdal66}
two-fold degeneracy, $\pi_{\e,\pm}$.  This has actually been done for
three cases, OGLE-2007-BLG-224 \citep{ob07224}, OGLE-2008-BLG-279
\citep{ob08279}, and MOA-2016-BLG-290 \citep{mb16290}.  The first two
were TP measurements, with all three observatories being on Earth.
The last was simultaneously observed from Earth and by the {\it Spitzer}
and {\it Kepler} satellites.  A crucial fact about this method
is that the third-observatory measurement can be of much lower quality
than the other two because it does not need to independently determine
the parameters of the event, but only break the degeneracy between the
implications from the two other measurements.  For example, in the case
of OGLE-2007-BLG-224, the degeneracy-breaking measurements from the Liverpool
Telescope (Canaries) contained only points $\ga 2\,$mag below the peak.
Nevertheless, when combined with knowledge of $t_\e$ from other
observatories, these determined $t_{0,\rm Canaries}$ to a precision of $<3\,$s,
even though these data, by themselves, contained essentially no information
about $u_{0,\rm Canaries}$.

Taking note of this fact, one could augment the system of two
GEO observatories (with $180^\circ$ relative phase) by a much less
capable observatory (or set of observatories with near-continuous
capability) on Earth.  For simplicity, consider the idealized situation
that the Earth observations take place from Earth's center.  Then, if
we label the projected positions of the three observatories
${\bf A}_\perp$, ${\bf B}_\perp$, and ${\bf C}_\perp$ (with Earth
being the last), we obtain ${\bf C}_\perp = ({\bf A}_\perp+{\bf B}_\perp)/2$.
And therefore $t_{0,\bf C} = (t_{0,\bf A} + t_{0,\bf B})/2$
and $u_{0,\bf C} = (u_{0,\bf A} + u_{0,\bf B})/2$.  The two degenerate solutions
yield the same values of $t_0$, so the first of these two equations
is not useful.  However, the second can make very different predictions
for $u_0$.  In the present case $|u_{0,-,\bf C}| = 0.025$, while
$|u_{0,+,\bf C}| = 0.375$.  Note that allowing the Earth observatory to
operate from its surface (rather than its center) changes the mathematical
details of this example, but not the basic principle.

\section{{LEO-only Observatory}
  \label{sec:leo}}

The manner in which the LEO observatory can break the degeneracy
between the two ($\pi_{\e,\pm}$) solutions derived from Earth+GEO
observations (Section~\ref{sec:degen}), raises the possibility that
LEO-only observations could measure $\pi_\e$ by itself, i.e., with no
help from a second observatory \citep{mogavero16}.  Clearly, the domain of such
measurements would be much more restricted than the GEO option that is
illustrated in Figure~\ref{fig:par2}.  For example, as just recounted
in Section~\ref{sec:degen}, the Mercury-mass FFP at $D_L\sim 5\,\kpc$.
would have an unmeasurably small parallax signal from LEO, whereas
it would be easily detectable using Earth+GEO.  On the other hand,
the 2-Pluto-mass FFP at $D_L\sim 1\,\kpc$ could be characterized from
LEO alone.

\begin{figure}
\plotone{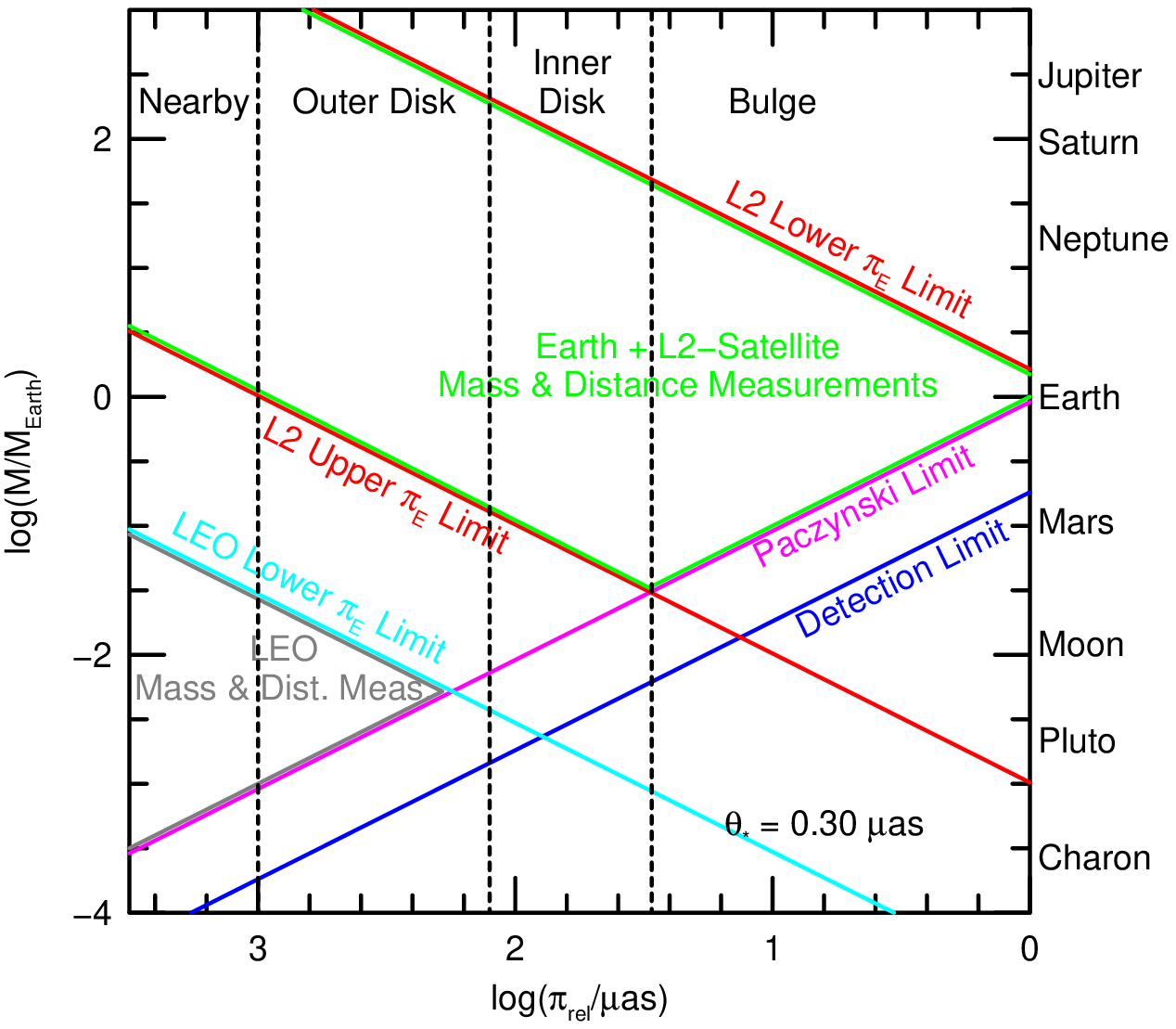}
\caption{Rough representation of the $M$-$\pi_\rel$ space that is
  accessible to FFP parallax measurements from LEO.  The Figure is
  constructed simply by substituting $D_\perp = a_{\rm LEO}=1\,R_\oplus$ for
  $D_\perp = a_{\rm GEO}=6.6\,R_\oplus$ in Figure~\ref{fig:par2}.
  Comparing to that Earth-GEO Figure, the LEO-only observations
  reach to even lower FFP masses but are restricted to even more
  nearby populations.  As discussed in Section~\ref{sec:leo}, several
  aspects of this Figure require deeper investigation based on simulations,
  so it should be regarded as a very rough guide to what is possible.
    }
\label{fig:par4}
\end{figure}

However, the overall prospects for such a LEO observatory cannot be
properly evaluated using the analytic techniques of the present work:
there are several interrelated issues that can only be addressed using
simulations.  One such issue has already been described, namely the effect
of the windowing of the observations by Earth occultations.  To help
understand the other issues, I present in Figure~\ref{fig:par4},
a LEO-style version of Figure~\ref{fig:par2}, which is calculated by
replacing $D_\perp=a_{\rm GEO}$ with $D_\perp=a_{\rm LEO}\rightarrow R_\oplus$.
This substitution is not strictly kosher, exactly because the resolution
of the issues raised in this section could alter some features of this
diagram and/or the interpretation of these features.  Nevertheless,
the Figure is useful because it permits us to frame the discussion of
these issues.

The second issue is the rapidly changing $\mu_\rel$ distribution for
lens distances $D_L\la 1\,\kpc$.  For distances $D_L\ga 2\,\kpc$, the
scale of the proper motion distribution is set by the ratio of
Galactic $v_\rot/R_0=\mu_{\rm Sgr A*}=6.4\,\masyr$.  But at nearby
distances, $D_L\la 2\,\kpc$, the local transverse stellar velocities
$\sigma_\perp\sim 37\,\kms$ divided by the distance, $\mu_{\rm loc}=
\sigma_\perp/D_L= 7.8/(D_L/\kpc)\,\masyr$, must be added in
quadrature.  These higher proper motions for nearby lenses are
relevant because, formally, the sensitivity of the LEO approach is
restricted to $D_L<3.2\,\kpc$, and only opens up to a dex of mass
range for $D_L<1.3\,\kpc$.  See Figure~\ref{fig:par4}.
As mentioned above, the sensitivity of
LEO observations to parallax is greatly affected by whether or not
significant magnification persists on both sides of an occultation,
or preferably, for at least one orbit, i.e., 90 min.  At the magenta
(``Paczy\'nski limit'') boundary, i.e., $\rho=2$, the significantly
magnified region does not extend much beyond the face of the source,
whose diameter in this illustration is $2\theta_*=0.6\,\mu$as.  Hence,
to be magnified for one orbit requires $\mu_\rel\la 2\theta_*/90\,\min =
3.5\,\masyr$.  Such slow proper motions are relatively rare when
the characteristic proper motions are $\mu_\rel\sim 6\,\masyr$, but become
much rarer for the higher values at $D_L\la 1\,\kpc$.

The third issue (which goes in the opposite direction of the second),
is that for LEO, the true limit on the measurability of parallax may
be closer to the ``Detection limit'' (blue line) than the ``Paczy\'nski limit''
(magenta line).  As discussed by \citet{gould24}, the reason that $\rho\gg 1$
events present difficulties for the two-observatory \citep{refsdal66}
parallax method is that the Paczy\'nski parameters must be separately
measured from each site, and \citet{johnson22} showed that this is
can be very difficult for $\rho\gg 1$.  However, parallax measurements
that rely on the acceleration of a single site, such as AP \citep{gould92},
GEO \citep{gould13}, and LEO \citep{honma99}, might not face an analogous
restriction.  This is particularly the case in the region between
the magenta and blue lines (and for nearby lenses) because $\pi_\e$
can be very large, leading to very large deviations from rectilinear
motion as seen from LEO.

For all these reasons, the efficacy of LEO-only parallax measurements
must be addressed by simulations.  These simulations will determine
whether {\it CSST}, which is already selected for launch, can be usefully
employed for FFP parallaxes, but they can also guide the design of new
LEO missions that have FFP parallaxes as one of their mission goals.

\section{{Discussion}
  \label{sec:discuss}}

To my knowledge, there is only one survey mission that is already planned
for either GEO or LEO, namely {\it CSST}.  Because this is a LEO mission,
no complementary ground-based observations are required to obtain parallaxes.
Hence, the only issue is to determine whether observations based on the
existing design will yield good FFP $\bpi_\e$ measurements.

However, for future LEO and GEO missions that have FFP parallaxes as
one of their mission goals, the design issues extend much more broadly
than have been explored in this work.  As one example, GEO is 40 times
closer to Earth's surface than L2, meaning that 1600 times less bandwidth
is required for the same data download.  This opens the possibility of
using vastly larger detector arrays relative to the L2 surveys, {\it Roman}
and {\it Euclid}, which in turn might require using optical CCDs rather
than infrared arrays, simply due to cost.  Related to this is the issue
of telescope aperture.  For fixed focal design, the sky area scales
inversely with the aperture, so that (ignoring slew/readout time and
sky/field-star background) one can reach the same signal-to-noise ratio
by observing a single field using a smaller aperture or tiling through
several smaller fields with a larger aperture.  The two caveats just
mentioned (readout and sky) work in opposite directions, and so plausibly
roughly cancel.  However, smaller-aperture telescope are vastly cheaper,
and (given the less severe bandwidth requirements just mentioned) one might
consider putting several such smaller telescopes on the same spacecraft
bus, in analogy to the approach pioneered by ground-based transit
surveys.

It is not my intention to explore all such issues in the present
work.  I merely point out that the problem FFP parallaxes in the
dwarf planet regime should be thought through from the bottom up.

\acknowledgments
I thank Subo Dong for valuable comments and suggestions.

\clearpage

\end{document}